\documentclass[twocolumn,draft,prb]{revtex4-1}

\usepackage{graphicx}
\usepackage{dcolumn}
\usepackage{bm}
\usepackage{subfigure}
\usepackage{dsfont}
\usepackage{xspace}
\usepackage[USenglish]{babel}
\usepackage{amssymb,amsmath}
\usepackage{mathtools}
\usepackage{relsize}
\usepackage{lmodern}
\usepackage{slantsc}
\usepackage{scalefnt}
\usepackage{color}
\usepackage{array}
\usepackage{ntheorem}
\allowdisplaybreaks[1]

\newcommand{\be}{\begin{equation}}
\newcommand{\ee}{\end{equation}}
\newcommand{\bse}{\begin{subequations}}
\newcommand{\ese}{\end{subequations}}
\newcommand{\bpm}{\begin{pmatrix*}}
\newcommand{\epm}{\end{pmatrix*}}
\newcommand{\bmm}{\begin{matrix}}
\newcommand{\emm}{\end{matrix}}
\newcommand{\sss}{\sigma}
\newcommand{\vac}{\vert \Phi \rangle}
\newcommand{\mU}{\mathcal{U}}
\newcommand{\mA}{\mathcal{A}}

\newcommand{\Z}{\mathbb{Z}}

\newcommand{\A}{\mathcal{A}}
\newcommand{\T}{\mathcal{T}}
\newcommand{\U}{\mathcal{U}}

\newcommand{\ket}[1]{|{#1}\rangle}

\newcommand{\ii}{\mathrm{i}}

\makeatletter
\newcommand*{\Relbarfill@}{\arrowfill@\Relbar\Relbar\Relbar}
\newcommand*{\xeq}[2][]{\ext@arrow 0055\Relbarfill@{#1}{#2}}
\makeatother

\newcommand{\lv}{\mathbf{l}}

\newcommand{\km}{$K$ matrix\xspace}
\newcommand{\kms}{$K$ matrices\xspace}

\newcommand{\1}{\mathbf{1}} 

\usepackage{boxedminipage}

\begin{document}


\title{Symmetry Enriched Phases via Pseudo Anyon-Condensation}

\author{Ling-Yan Hung}
\email{jhung@perimeterinstitute.ca}
\affiliation{Department of Physics, Harvard University, Cambridge MA 02138, USA}
\author{Yidun Wan}
\email{ywan@meso.t.u-tokyo.ac.jp}
\affiliation{Perimeter Institute for Theoretical Physics, Waterloo, ON N2L 2Y5, Canada}

\date{\today}

\begin{abstract}
We show that a large class of symmetry enriched (topological) phases of matter in $2+1$ dimensions can be embedded in ``larger'' topological phases --  phases describable by larger hidden Hopf symmetries. Such an embedding is analogous to anyon condensation, although no physical condensation actually occurs. This generalizes the Laudan-Ginzburg paradigm of symmetry breaking from continuous groups to quantum groups---in fact algebras---and offers a potential classification of the symmetry enriched (topological) phases thus obtained, including symmetry protected trivial phases as well, in a unified framework.

\end{abstract}

\maketitle
%

Recently, there is a surging progress\cite{Chen2011f,Senthil2012a,Vishwanath2012,Essin2012,Mesaros2011,Hung2012a,Wang2012,Lu2013,Levin,HungWan2013a,Metlitski2013,Ye2013,Sule2013,Xu2013} in constructing and classifying gapped phases with symmetries, namely the symmetry enriched phases (SEPs), including symmetry protected trivial (SPT) phases and symmetry enriched topological (SET) phases. Understanding these phases is a fundamental question in condensed matter physics which also promises potential applications, such as topological quantum computation. In this letter, we propose a systematic construction of SEPs by embedding them in different intrinsic topological orders. We recover the  SEPs via \emph{condensation of anyons}, although here we only borrow the Mathematics, and no physical condensation actually occurs. The Mathematics here is in fact that of Hopf symmetry breaking, as generally a topological order has a hidden symmetry algebra, a Hopf algebra, which is broken by the anyon condensate. This thus generalizes Landau's symmetry breaking paradigm.

The Hopf breaking picture reproduces all the 2+1 $d$ SET phases thus far constructed via the projective symmetry group (PSG).
As such, to understand the Hopf breaking, we begin with the PSG, treating it as a special case. Consider an SET phase $S$, a deconfined gauge theory with gauge group  $N_g$ and a global symmetry $G_s$; for later convenience, we call the topological order described by the gauge group $N_g$ the {\bf bare phase} of $S$. The phase $S$ is specified by assigning compatible $G_s$ charges, potentially fractionalized, to the $N_g$ gauge charges (chargeons), fluxes (fluxons) and dyons, which are generally anyons.  A PSG group $G$ can encode such an assignment, such that $N_g$ is its normal subgroup, and that $G_s = G/N_g$.  The embedding of $N_g$ in $G$ determines the non-trivial $G_s$ representations the anyons fall into.

An alternative way exists, however, to encode the quotient group structure, by choosing a ``charge condensate'' $\vac$ such that $N_g$ is the invariant subgroup of $G$ that keeps $\ket{\Phi}$ invariant. This is familiar in the Higgs mechanism, where the quotient $G_s= G/N_g$ describes the moduli space, and in the PSG construction $G_s$ is interpreted as the global symmetry. The condensate $\vac$ by definition transforms under $G_s$ and thus carries a $G_s$ charge, but it carries no $N_g$ charge and thus automatically belongs to the topologically trivial sector of the $N_g$ gauge theory. To emphasize the distinction from a physical condensate, let us call $\vac$ the {\bf charged vacuum}.

In the context of a topological order, the chargeons and fluxons of the deconfined gauge theory are on equal footing. The ``anyon condensate'' perspective of the PSG construction therefore immediately suggests a generalization by considering more general $\vac$.

We therefore need to go beyond gauge symmetry breaking. It turns out Hopf symmetry breaking, which naturally treats chargeons, fluxons and dyons equally, can describe general anyon condensates. In fact, for a general topological phase not necessarily related to a gauge theory, each type of anyon is simply a representation of some given hidden Hopf symmetry algebra.  Hopf symmetry breaking has been discussed in the context of anyon condensation\cite{Bais2002,Bais2003,Bais2009,Bais2009a,Bais2012,You2013}, where a macroscopic condensate of anyons appears, driving a phase transition that takes one topological phase described by Hopf algebra $\mathcal{A}$ to another phase described by $\mathcal{U}$. Here, the \emph{same} Mathematics encodes the physics of SEPs. A topological phase with a Hopf symmetry $\A$, which plays the role of PSG, encodes an SEP phase whose bare phase has a smaller Hopf symmetry $\mathcal{U}$.
Let us summarize the four steps of how this is to be done:

\textbf{Step 1: The charged vacuum $\vert \Phi \rangle$}.
As discussed above, we first choose a $\vac$ in the representations of $\mathcal{A}$. In the PSG construction $\vac$ is just a state in a representation of the PSG $G$.
The topological order in the SEP phase $S$ is then related to the subalgebra of $\mA$, denoted $\mathcal{T}$, that keeps $\vert\Phi\rangle$ ``\emph{invariant}'', a subtle concept that has to be defined carefully in the context of an algebra,  and such an approach has indeed been taken in Ref\cite{Bais2003}.  This approach is well defined in Hopf algebras such as the twisted quantum double (TQD)\cite{Hu2012a,Mesaros2011}, where the internal states of all representations are physical and appear in the physical Hilbert space. In these cases however, the quantum dimension of the anyon (representation) containing the "charged vacuum" is necessarily an integer.

More generally, anyons have fractional or even irrational quantum dimensions, e.g., the anyons described by quantum groups $U_q(sl_2)$ for $q$ a root of unity; nonetheless, it is still possible to define $\vert \Phi \rangle$ as hidden in a representation of $\mA$\cite{Bais2009a}, although one does not solve for the subalgebra $\mathcal{T}$ explicitly. Instead, one assumes\footnote{Here we actually assume that the anyons are described by a unitary braided tensor category, which is the case for twisted quantum doubles and quantum groups. This assumption enforces that an anyon $a$ and its dual $a^*$ annihilate to the vacuum in a unique way, namely $a\times a^*=1+x$, where for simplicity $x$ collects all other anyon content that does not contain the vacuum. Therefore a sector $a$ that has quantum dimension larger than one potentially has to split after anyon condensation to preserve unitarity.} that each representation (anyon) $a$ of $\mA$ decomposes to representations (anyons) $a_i$ of $\mathcal{T:}$
\be \label{eq:restrict}
a = \sum\nolimits_i n^{a}_i  a_i,
\ee
where $n^{a}_i$ is the multiplicity of the representation $a_i$ of $\mathcal{T}$ in the decomposition. Note that the decomposition conserves the quantum dimension, $d_{a} = \sum\nolimits_j n^{a}_i d_{a_i}$.

One can choose $\vac$ to be one (or more) of these $a_i$'s hidden in an (or several different) $a$. The allowed quantum dimension for $\vac$ must be one, as expected of a physical excitation, and the \emph{topological spin} $\theta_{\vac}$ of $\ket{\Phi}$, the self statistics of $\vac,$ should be unity. Prior discussions of SEP phases describable by gauge theories enriched by global symmetries\cite{Hung2012a} call $\vac$  a ``pure ($G_s$) charge', which has bosonic self statistics and trivial mutual statistics with all other electrically gauge charged particles,  in accord with our discussion.

\textbf{Step 2: Identifying the ``bare phase'' and the confined sector}. One might expect the ``bare phase'' to be described by the ``invariant subalgebra'' $\mathcal{T}$ above, but this is not true\footnote{We assume the quasi-triangular Hopf algebra $\A$ is modular, then there is always a confined sector in $\T$, such that $\U$ is smaller than $\T$}. According to the literature on anyon condensation, an anyon condensate always divides the the anyons into two distinct sets: the confined and unconfined sectors. The former consists of all the anyons that are \emph{non-local} with respect to the condensate, whereas the latter consists of those local with respect to the condensate. Two excitations are mutually local if they have trivial mutual statistics. An anyon mutually non-local with the condensate would pull a string in the medium, and its creation or isolation is thus energetically expensive, i.e., they are indeed \emph{confined}. This concept is again applicable here in determining the ``bare phase''.  Recall in Step 1 that $\vac$ is taken as the charged vacuum. Therefore, barring the $G_s$ charge it carries, as far as the ``bare phase'' is concerned, it behaves exactly as the trivial sector, which is necessarily mutually local with \emph{all} other anyons of the ``bare phase''. As such, the anyons in the confined sector cannot be part of the ``bare phase''.  Discussed in Ref\cite{Lu2013,Hung2012a,Wen2013a}, confined particles are precisely those identified as the ``twist'' particles.  As a result, quite generally, the  Hopf algebra $\U$ that characterizes the ``bare phase'' differ from the ``invariant subalgebra'' $\T$ introduced above, as some of $\T$'s anyon representations are excluded from $\U$.

To recover $\U$, we should first isolate the confined sector. For this we need the fusion algebra of the decomposed anyons $a_i$ and $b_j$, which we can deduce from two facts: 1) the fusion rules commute with (\ref{eq:restrict}), namely
\be \label{eq:commute}
a\otimes b = c \to (\sum\nolimits_{i,j} n^a_i n^b_j a_i \otimes b_j )= \sum\nolimits_k n^c_k c_k ,
\ee
which implies the conservation of quantum dimension,
and 2) the fusion algebra is associative.
Consider particularly the fusion of the decomposed pieces $a_i$ and $\vac$. Since $\vac$ is topologically trivial in $\U$,  any two anyons belong to the same topological sector in $\U$ if they are related by fusion with $\vac$. But if two supposedly identified anyons $a_i$ and $b_j$ descend from anyons $a$ and $b$ in $\A$ with topological spins $\theta_{a} \neq \theta_{b}$, then as anyons in $\U,$ they cease to have well defined topological spins; hence, they must be confined. According to Ref\cite{Bais2009a}, this is an if-and-only-if condition for determining the confined anyons.

With all confined anyons excluded and all anyons related by $\vac$ identified, what remains is the unconfined sector consisting of anyons with well defined topological spins and fusion rules between them, leading to a unique $\U$. Clearly to describe SET phases, the resultant $\U$ is non-trivial with multiple topological sectors, whereas a trivial $\U$ encodes the physics of SPT phases. We note that in practice there are further consistency conditions that has to be taken into account to obtain the complete fusion algebra, as detailed in \cite{Bais2009a}.

\textbf{Step 3: Identifying the global symmetry group action and the pure braids}. The story of anyon condensation ends, but our journey has only just begun: we need to find out the global symmetry $G_s$ and how it transforms the anyons in $\U$. The key lies in the confined sector. Let us  propose a rule for Abelian $\A$ and $\U$:

\centerline{\begin{boxedminipage}[c]{0.98\columnwidth}One can extract the group action of an element $g_i$ of $G_s$ by winding the physical system (a set of particles in $\mU$) around a \textit{spectator}, a specific confined particle $c_i$ corresponding to $g_i$. The fusion of the spectators modulo the fusion of the unconfined sector yields the structure of $G_s$. That is, the confined sector generates the group action of the non-trivial elements of $G_s$; the unconfined particles obtainable from fusing the confined particles generate the action of the identity, and are thus related to global charge fractionalization.\end{boxedminipage}}

The gauging procedure in Ref\cite{Levin2012,Hung2012,Hung2012a,Lu2013} inspires the above proposal. When the global symmetry is gauged, the new theory has an expanded spectrum containing excitations carrying ``magnetic flux'' of the global symmetry $G_s$. In those cases studied, the Aharanov-Bohm phase acquired by a  $G_s$ charge moving about a $G_s$ flux coincides with a particular group action of $G_s$, depending on the flux particle involved.

This can be contrasted with Ref\cite{Essin2012} which gives a consistency condition of the $G_s$ action, that the identity element of $G_s$ must act on individual anyons in a way such that its aggregate action on a system of anyons fusing to a physical state is trivial. This condition strongly points to identifying confined particles as ``generators of $G_s$'', up to their fusion to unconfined particles. This also complies with our intuition from group symmetry breaking, where $G_s$ is certain quotient of $\mA$ by $\mU$.

A complete generalization of the above statements to non-Abelian $\mA$ and $\mU$ is unclear.
 For example, quite generally introducing an extra confined anyon into the system of unconfined anyons enlarges the Hilbert space. Winding the physical system around the confined particle is given by some general unitary operator that acts on the enlarged Hilbert space. It is thus not immediately obvious whether we can systematically truncate the operator to act on the subspace corresponding to one that describes the original physical system. Also the fusion of confined particles generally lead to a direct sum of anyons, potentially containing both confined and unconfined ones, rendering the $G_s$ group structure unclear.

There are however large classes of special cases where some statements can be made. When $\mA$ is a TQD of $G$, the anyons carry integer quantum dimensions, as they are basically irreducible representations of the group $G$ and its various subgroups. The braiding of anyons always factorizes into actions on the internal spaces of each anyon present. Here, one can also read off the group action of $G_s$ by braiding.  The fusion algebra of the confined particles there is consistent with the group product in the sense that the Aharanov-Bohm phase acquired through braiding with a set of confined particles is equal to the product of these transfomations obtained from braiding with individual constituents in the set of confined particle in a specific order that can be understood from the property of the fundamental group governing the combining of loops\cite{Preskill2004}.
A non-Abelian $\Z^3_2$ twisted gauge theory can be broken down into a novel SPT phase that involves rotation of different species, to be reported in a forth-coming paper\footnote{L.Y. Hung, Y. Wan, to appear}. An ambiguity arises in the overall phase factor if the confined particle has quantum dimension $>1$ as it also suffers a transformation of its internal space upon braiding with an unconfined particle.

For more general non-Abelian $\A$, well known examples also exist where the fusion algebra of the confined particle reproduces the group structure $G_s$. Consider  $\A$ corresponding to the Ising$\times \overline{\textrm{Ising}}$ theory. An SET is embedded in which the $\Z_2$ toric code is endowned with a $\Z_2$ global symmetry, which exchanges the $e$ and $m$ excitations\cite{Lu2013,HungWan2013a}. The fusion of any two of the four confined anyons $\{(\sigma,1), (1,\sigma), (\psi,\sigma), (\sigma, \psi)\}$ again yields an unconfined particle, which indeed reflects the $\Z_2$ structure. This is studied in the supplemental material.

In any event, the above proposal reproduces all the known SEPs and their relationship with the gauged theory, barring the precise relation between confined particles and $G_s$ structure. It also opens a door to generating new phases beyond the PSG construction and the \km formalism, the latter method is suitable only for Abelian ``bare phases''.

In most cases in this letter, the anyons fall into the previously defined three kinds: chargeons, fluxons, and dyons, all having integer quantum dimensions and can be chosen as the condensate. For Abelian anyons, the corresponding LRE phase admits a simple formulation as a Chern-Simons (CS) gauge theory in terms of \kms,
\be\label{eq:KmLagrangian}
L_{CS} = K_{IJ} a^I \wedge da^J/4\pi,
\ee
with $U(1)$ gauge fields $a^I$. To be specific, consider $G=\Z_n$, the \km is
\be\label{eq:Z4Kmatrix}
K=\bpm
2m & n\\
n & 0\epm,
\ee
where $m\in\Z_n$ labels the $n$ possible phases described by \eqref{eq:KmLagrangian}. The anyons then fall into vectors $\lv=(l_1,l_2)\in\Z_n\times\Z_n$, where $l_1$ and $l_2$ label respectively the $\Z_n$ charge and flux. The mutual statistics between two anyons $\lv^a$ and $\lv^b$ is $\theta_{ab}=2\pi\lv^a\cdot K^{-1}\cdot \lv^b$. Vector addition modulo the trivial sector readily yields the fusion rules\cite{Lu2013,HungWan2013a}. Now we explore an explicit example in this simple situation before heading to Step 4.

\textbf{Example 1: $n=4$ and $m=0$ dyonic breaking}. Here we let $\lv^{\ket{\Phi}}=(2,2)$. The anyons' mutual statistics demands that the unconfined anyons $\lv^u$ must satisfy
\be\label{eq:N4m0unconfined}
l^u_1+l^u_2\in 2\Z.
\ee
Having identified the anyons related by fusing with $\lv^{\ket{\Phi}}$ and/or $\lv^B=(4p,4q)$ with $p,q\in\Z$, the trivial particles in the $\Z_4$ gauge theory, we know the distinct $\U$ anyons are
\be\label{eq:N4m0Uanyons}
\lv^u=(2t-s,s),\quad t,s\in\{0,1\}.
\ee
The self statistics of  such an $\lv^u$ reads $\theta_u=\pi s(t-s/2)$, which is semionic. This means $\U$ describes a doubled semion model. The mutual statistics also enables one to parameterize the confined sector by $\l^c=(l^c_1,l^c_2)$ with $l^c_1+l^c_2$ odd. Thus clearly, any two confined anyons fuse to an unconfined one, implying that $G_s=\Z_2$. As such, we obtain from the topological phase described by a $\Z_4$ gauge theory an SET phase: the doubled semion model enriched by $\Z_2$ global symmetry. To see how this $\Z_2$ symmetry transforms the $\U$ anyons, following the Step 3 introduced above, one picks a spectator $\lv^q$ from the confined; here, for simplicity we take $\lv^q=(1,0)$. Note that any confined in this example works the same. The $G_s$ charge of an unconfined anyon reads $Q_u=\lv^q\cdot K^{-1}\cdot\lv^u$, which readily gives rise to Table \ref{Z4tb3} in the supplemental material. From electric-magnetic duality it is clear that pure flux breaking is identical to the PSG construction.

\textbf{Remarks}: In the case with $n=4$ and $m=2$, condensing $\lv^{\ket{\Phi}}=(1,2)$ simply produces the bosonic SPT phase with $\Z_4$ global symmetry, by a procedure similar to that in Example 1. On the other hand, the electric breaking for any $m$ value in \eqref{eq:Z4Kmatrix} via condensing $l^{\ket{\Phi}}=(2,0)$ reproduces the PSG-constructed SET phases\cite{Hung2012a}.

{\bf Step 4: Edge states}. One may introduce a boundary to an SEP and study the edge modes.
An important question regards the appearance of robust gapless modes. Non-trivial SPT phases have non-trivial edges, while a non-trivial SET phase may have a trivial edge, one that is gapped without breaking the global symmetry\cite{Wang2012,Lu2013,HungWan2013a}.

To begin with, let us consider the case where the gauged SET described by $\mathcal{A}$ is in fact Abelian, and the necessary and sufficient condition for a trivially gapped edge for Abelian topological order has been given\cite{Levin}. The condition boils down to the existence of a Lagrangian subset $\mathcal{L}$, comprising topological sectors that have trivial mutual statistics, and that any topological sector not in $\mathcal{L}$ must have non-trivial mutual statistics with at least one member of $\mathcal{L}$. Note that $\mathcal{L}$ is closed under fusion.

As encountered in Example 1, we can describe an Abelian $\A$ as a CS theory, characterized by an $N\times N$ symmetric $K_\A$ matrix, and whose quasi-particles are characterized by $N-$component integer vectors. Again we restrict to non-chiral phases, where $K_\A$ has an equal number of positive and negative eigenvalues, implying that $N$ is even.

The bare phase $\U$ of the Abelian SET is characterized by another matrix $K_\U$. The anyon condensation process that connects  $K_\U$ and $K_\A$ is understood as follows\cite{Lu2013}:
\be
K_\A^{-1} = M^T K_\U^{-1} M,
\ee
and that excitations in $\A$ and those of $\U$ are related by
\be
l_\U = M l_\A,
\ee
where the matrix $M$ ensures that all the flux particles generating global symmetry on the SET phase are valid quasi-particles in the theory of $K_\A$.
This relation between the SET and the gauged theory ensures that the gauging preserves the braiding/commutation properties between quasi-particles.
The above transformation guarantees that every boson (in the trivial sector) in $K_\A$ is automatically a boson in $K_\U$. In particular, this implies that every bosonic excitation belonging to the self and mutually \emph{commuting} set (i.e. a set of $\{\lv_{i}\}$ satisfying $\lv_i\cdot K_{\A}^{-1}\cdot \lv_j=0$) also correspond to commuting bosons in $K_\U$, and we recall that condensation of these bosons are responsible for gapping the edge.

But the converse of the above statement is not true: Some bosons, and also self-commuting bosons in $K_\U$ correspond to non-trivial topological sectors in $K_\A$. These non-trivial topological sectors are precisely those corresponding to the charged vacuum $\lv^{\vac}$.
Suppose at least one Lagrangian subset exists for the phase $\mA$, such that its edge is gapped. The $\mathcal{L}_{\A}$ ensures that there is a \emph{complete} set of $N/2$ linearly independent mutually-commuting bosons that can condense simultaneously. The \textit{completeness} reads that any other condensable  boson should not commute with at least one member of the above set.   Now, does the set of condensable bosons always stay \emph{complete} from the point of view of $K_\U$?

The answer is ``No". First, members of an $\mathcal{L}$
can be represented by a set of $l$-vectors arranged to be self-null for all $i,j \in \mathcal{L}$. Second, integer combinations of these $l$-vectors can reproduce the entire set of condensable bosons\cite{Wang2012,Barkeshli} . Hence, condensation of the complete set of condensable bosons automatically leads to the condensation of the entire $\mathcal{L}$.
So a complete condensable set in $K_\A$ potentially does not map to a complete condensable set in $K_\U$, and the missing bosons correspond precisely to the $\lv^{\ket{\Phi}}$. This scenario arises when $\lv^{\vac}\in \mathcal{L}$ characterizing the condensed, complete set of bosons in $K_\A$. Physically, since $\lv^{\vac}$ carries a global symmetry charge, were it part of the Lagrangian subset, which physically condenses at the boundary, the edge would break the corresponding global symmetry. Summarising, the edge of an SET can be gapped without breaking $G_s$ if there exist a Lagrangian subset in $K_\A$ which does not contain $\vac$.

{\bf Remarks on Non-Abelian $\mA$:}
In non-Abelian phases, a general understanding of gapping conditions is so far lacking. Note that one can rephrase the property of the Lagrangian subset as follows in terms of anyon condensation. {\bf A Lagrangian subset is a collection of anyons such that they can condense at the same time and that the resultant phase is a trivial order with only a single topological sector.}
The above statement however admits a direct generalization to non-Abelian phases; therefore, it is tempting to conjecture that it is the gapping condition for non-chiral theories where topological central charges $c_L= c_R$\cite{Kitaev2006} . This conjecture would suggest that the corresponding gapping condition of the SET with topological order $\mU$ is such that the charged vacuum $\vac$ belongs to the generalized Lagrangian subset $\mathcal{L}$ as defined above.

{\bf Fractionalization?} Before we conclude, let us comment on the allowed ``global charge fractionalization'' of a given bare phase $\mU$.  As aforementioned, pure braids between two unconfined anyons should generate consistent actions of the \emph{identity} element of $G_s$. Such an action on an anyon uniquely fixes the anyon's fractional charge. We can thus read off the allowed fractional charges from braiding properties of the bare phase regardless of $G_s$, and how that is encoded in $\mA$. That is, consistent fractionalization is intrinsic to the bare phase. Consider one simple non-Abelian example, the Ising model. There are three topological sectors $1,\sigma,\psi$, whose topological spins are $h_1 = 1$, $h_\psi = \frac{1}{2}$, and $h_\sigma = \frac{1}{16}$. The non-trivial fusion rules are $\sigma \otimes \sigma = 1 \oplus \psi$ and $\sigma \times \psi = \sigma$. There are readily two possible fractionalization actions of any $G_s$:

\textbf{1. Identity generator by braiding with} $\psi$:
$\Phi(x)= \exp(2\pi i (h_{x\otimes \psi} - h_x - h_\psi)) $
gives the monodromy phase factor for each sector $x\in \{1,\sigma,\psi\}$. Explicitly, we have
\be
\Phi(1) =\Phi(\psi)=1,\qquad \Phi(\sigma) = -1.
\ee
This shows that 1 and $\psi$ transform as linear representations, but the global charge of $\sigma$ is fractionalized into $1/2$ and can transform projectively under any given $G_s$. the central extension/PSG characterizing the projective representation is classified by $H^2(G_s,\mathbb{Z}_2)$.

\textbf{2. Identity generator by braiding with} $\sigma$:
Consider a lump of $2N$ $\sigma$'s as our physical system, and one extra spectator $\sigma$ to be taken as our symmetry generator.  Consider the pure braid in which the spectator $\sigma$ encircles the physical system. The braiding matrix can be shown to be
$e^{\ii N \pi /2 }\prod_{i=1}^{2N} \gamma_i$, where $\gamma_i$ are Clifford gamma matrices\cite{Nayak2008}.
This operator measures the overall topological charge. For +1 eigenstates they correspond to overall fusion to $1$, and $ -1$ eigenstates to $\psi$. The braid matrix encodes a $\Z_2$ subalgebra of the fusion algebra, although this cannot be understood as a local action on each anyon, but an aggregate action on the Hilbert space.

{\bf Conclusion:} In this letter we present a generalization of the PSG construction of SEPs, by considering anyon condensation, and demonstrate how the physics of SEPs is encoded in a larger topological phase. Given recent progress in understanding anyon condensation\cite{Kitaev2012,Kong2013}, our methods is likely a first step toward a systematic classification of SEPs, particularly non-Abelian SETs.

\begin{acknowledgments}
We thank FA Bais, P Gao, ZC Gu, S Haaker, L Kong, M Levin, FL Lin, YM Lu, XL Qi, Y Ran, ZH Wang, XG Wen, and YS Wu for helpful discussions, and in particular Juven Wang for his meticulous proof reading of the manuscript. LYH is supported by the Croucher Fellowship. This research was supported in part by Perimeter Institute for
Theoretical Physics. Research at Perimeter Institute is
supported by the Government of Canada through Industry
Canada and by the Province of Ontario through the
Ministry of Economic Development \& Innovation.
\end{acknowledgments}

\bibliographystyle{unsrtnat}
\bibliography{StringNet}

\newpage
\section*{Supplementary Material}


\subsection{Dyonic breaking of $\Z_4$ gauge theory}
We describe an example in which a $\Z_4$ gauge theory is broken down
into a doubled semion model with global $\Z_2$ symmetry. We summarise
the results below in Table \ref{Z4tb3} that can be readily compared to results in \cite{Hung2012a}. It corresponds to the phase labeled $(011)$ in \cite{Hung2012a}. It is remarkable that here that the dyonic breaking leads to the same result as electric breaking of a \emph{different} gauge group, namely $\Z_2\times \Z_2$.
Note that particles with a unit of ``$G_s$ twists'' are by definition confined particles.

\begin{table}[h!]
\centering
 \begin{tabular}{ |c||c|c|c| c|}
 \hline
 \multicolumn{5}{|c|}{$m=0$, $l^{\vac} =(2,2) $} \\
 \hline
                &$(l_1l_2)$ & $G_s$-charge & $G_s$-twist & statistics \\
 \hline
                &(00) & 0 & 0 & 0\\
                &(22) & 1 & 0 & 0\\
                &(02) & 1 & 0 & 0\\
Unconfined&(20) & 0 & 0 & 0\\
                &(11) & 1/2&0& 1/2\\
                &(13) & $-1/2$&0&$-1/2$\\
                &(31) & $1/2$&0& $-1/2$\\
                &(33) & $-1/2$&0&$1/2$\\
 \hline
                &(01) & $1/2$ & 1 & 0\\
                &(10) & 0 &1 & 0\\
                &(03) & $-1/2$&1&0\\
Confined&(30) &0&1 &0\\
                &(21) &$1/2$&1& 1\\
                &(12) & 1 & 1& 1\\
                &(32) & 1&1&1\\
                &(23)&$-1/2$&1&1\\
\hline
 \end{tabular}
\caption{
 The  $G_s$-charges, the  $G_s$-twists,
 the $G_g$-gauge sectors, and the statistics of the 16 kinds of
 quasiparticles/defects in the SET state $m=0$ with $l^q =(1,0)$.
}
\label{Z4tb3}
\end{table}

\subsection{Unconventional on-site transformation and Non-Abelian topological phases}
The above examples focus on symmetry breaking in which the broken theory $\mA$ is Abelian.
While our understanding of non-Abelian $\mA$ is incomplete, we would like to demonstrate
here an example illustrating how our discussion thus far can be generalized.

\subsubsection{The $Z_2$ symmetric toric code model and Ising${}^2$ model}
This example is inspired by the explicit example of anyon condensation in which an Ising $\times {\overline{\textrm{ising}}}$ theory gets broken down to a $\Z_2$ toric code, and the observation in \cite{Bombin2010, Kitaev2012} that some dislocations in the Kitaev toric code model, such that the $e$ and $m$ excitations are exchanged upon crossing it, satisfies the fusion that resembles the Ising model.  We will demonstrate how the framework proposed in this paper suggests that the $\Z_2$ symmetry enriched toric code in which $e,m$ are interchanged under the symmetry action is embedded in the Ising $\times {\overline{\textrm{Ising}}}$ theory.

Consider therefore $\mA$ that describes the Ising $\times {\overline{\textrm{Ising}}}$ model.
This is a direct product of two Ising, such that each copy has three topological sectors $1,\sigma,\psi$ with topological spins given by $h_1= 1, h_\sigma= 1/16$, and $h_\psi = 1/2$, and quantum dimensions $d_1=1$, $d_\psi = 1$, and $d_\sigma= \sqrt{2}$ respectively. The fusion rule is
already reviewed earlier.
There are thus altogether nine topological sectors, each denoted by a pair $(x,y)$, $x,y \in \{1,\sigma,\psi\}$. The topological spin of each sector is then $h_{(x,y)} = h_x- h_y$, where as the quantum dimensions are similarly given by $d_{(x,y)} = d_x d_y$. We note that the copy ${\overline{\textrm{ising}}}$ denotes that the topological spins have their signs reverted, leading to the $-$ sign in front of $h_y$.

The charged vacuum is chosen to be $\vac = (\psi,\psi)$, which indeed has self-statistics unity.
Now by checking the fusion with $\vac$, one can deduce $\mU$.
From
\be
(\vac\equiv 1) \otimes (\sigma,1) = (\sigma,\psi)
\ee
we immediately conclude that $(\sigma,1)$ and $(\sigma,\psi)$ belong to the confined sector since they have different topological spins, and similarly for $(1,\sigma)$ and $(\psi,\sigma)$.
We also note that the fusion rules among confined particles are exactly those observed in the fusion between dislocations in Ref\cite{Bombin2010}. The dislocations described there  correspond to our confined particles. Also the fusion between any two confined particles always lead to a direct sum that contains only unconfined particles, confirming the $\Z_2$ global symmetry structure. As already noted in Ref\cite{Bais2009}, the confined particles in an anyon condensate is related to the boundary conditions in the resultant topological order.

On the other hand, we have, within $\mU$,
\be
(1,\psi) \sim (\psi,1) = \epsilon, \qquad (\sigma,\sigma) = e \oplus m.
\ee
Also, from
\begin{eqnarray}
(\sss,\sss)\otimes (1,\psi) &&= (\sss,\sss),\\
(\sss,\sss) \otimes (\sss,\sss) &&= (1,1) \oplus (\psi,\psi) \oplus (1,\psi) \oplus (\psi,1) .
\end{eqnarray}
and using associativity and commutativity of fusion with decomposition, one can deduce that\cite{Note3}
\be
e^2= m^2 =\epsilon^2= 1,\qquad e m \sim \epsilon.
\ee
We thus identify $\U = D(\Z_2)$, which is the hidden Hopf symmetry of the $\Z_2$ toric code.
We note that the discussion here agrees qualitatively with a very similar observation in \cite{Lu2013} which relies on a construction based on the boundary CFT.

Now we are interested in the $G_s$ group action and we would like to consider braiding between the unconfined sector and the confined sector.  As mentioned above, in explicit constructions of the toric code models with dislocations\cite{Bombin2010,Kitaev2012}, one sees explicilty that pulling $e$ across a dislocation turns it into $m$ and vice versa, which is interpreted as the $\Z_2$ group action. Here we would like to pursue a second avenue to demonstrate this phenomenon, via using the Clifford algebra as the representation of the Ising model\cite{Nayak2008}. Let us consider the Hilbert space of the three anyon system:
$[(\sss,\sss), (\sss,\sss)] (1,\sss)$.

We can represent the Hilbert space as the vector space acted on by the direct product of $\gamma_i\otimes \Gamma_I$, where $i \in \{\1,2\}$ and $I \in \{1,2,3\}$.

The basis states are given by $v_1 \otimes v_2$, where $v_{1,2}$ are each two component vectors.
Monodromy between two $(\sss,\sss)$ particles at $i$ and $j$ is effected by the matrix
\be
M^{(\sss,\sss), (\sss,\sss)}_{ij} = \gamma_{ij} \otimes \Gamma_{ij}
\ee
whereas the monodromy between a $(\sss,\sss)$ and $(1,\sss)$ at position $i,j$
would be given by
\be
M^{(\sss,\sss), (1,\sss)}_{ij} =  e^{\pi i /4} 1 \otimes \Gamma_{ij}.
\ee

To obtain states that would correspond to $e^2 (1,\sss), m^2(1,\sss), em(1,\sss)$ and $me(1,\sss)$, we consider the transformation of these states under monodromies.

For $e^2(1,\sss)$ and $m^2(1,\sss)$ we expect that they are invariant under $M^{(\sss,\sss), (\sss,\sss)}_{12} $, whereas
 $em(1,\sss)$ and $me(1,\sss)$ should have eigenvalue $-1$. On the other hand, under $M^{(\sss,\sss), (1,\sss)}_{23}$,
we expect $e^2(1,\sss)\leftrightarrow em(1,\sss)$ and similarly $m^2(1,\sss)\leftrightarrow me(1,\sss)$ up to an overall phase, and finally
under $M^{(\sss,\sss), (1,\sss)}_{13}$ we also expect $e^2(1,\sss)\leftrightarrow me(1,\sss)$ and $m^2(1,\sss)\leftrightarrow em(1,\sss)$.

With these constraints, and using the basis $\Gamma_i= \sigma_i$ and similarly for $\gamma_i$, where
$\sigma_{1,2,3}= \sigma_{x,y,z}$ are the Pauli matrices, one can choose to make the following identifications:
\begin{eqnarray}
e^2(1,\sss) &&= \frac{1}{2} \left((1,0)\otimes (1,0) + (0,1) \otimes(0,1)  \right), \\
m^2(1,\sss) &&= \frac{1}{2} \left((1,0)\otimes (1,0) - (0,1) \otimes(0,1)  \right), \\
em (1,\sss) && = \frac{1}{2} \left((1,0)\otimes (0,1) + (0,1) \otimes(1,0)  \right), \\
me (1,\sss) && = \frac{1}{2} \left((1,0)\otimes (0,1) - (0,1) \otimes(1,0)  \right).
\end{eqnarray}

We have written the basis as eigenstates of $\gamma_1\gamma_2 \otimes \Gamma_1\Gamma_2$, which
therefore admits an interpretation as adopting basis in the fusion channels
$[(\sss,\sss)(\sss,\sss)](1,\sss)$.
We can also rewrite the states in terms of eigenvectors  of $\gamma_1\gamma_2 \otimes \Gamma_2\Gamma_3$, which admits an interpretation as basis states in the fusion channels
$(\sss,\sss)[(\sss,\sss)(1,\sss)]$, which gives
\begin{eqnarray}
e[e(1,\sss)] &&= \frac{1}{2} \left((1,1)\otimes (1,1) + (-1,1) \otimes(-1,1)  \right), \\
m [m(1,\sss)] &&= \frac{1}{2} \left((1,1)\otimes (1,-1) + (1,-1) \otimes(1,1)  \right), \\
e [m (1,\sss)] && = \frac{1}{2} \left((1,1)\otimes (1,1) + (-1,1) \otimes(1,0)  \right), \\
m [e (1,\sss)] && = \frac{1}{2} \left((1,-1)\otimes (1,1) + (1,1) \otimes(-1,1)  \right).
\end{eqnarray}
By rewriting the states in the above basis, an interesting interpretation emerges.
The above can be given an interpretation such that
\begin{equation}
e(1,\sss) = \frac{1}{\sqrt{2}}\left( [(\sss,\sss)(1,\sss)]_{(\sss,1)} -  [(\sss,\sss)(1,\sss)]_{(\sss,\psi)}\right)
\end{equation}
\begin{equation}
m(1,\sss) = \frac{1}{\sqrt{2}}\left( [(\sss,\sss)(1,\sss)]_{(\sss,1)} +  [(\sss,\sss)(1,\sss)]_{(\sss,\psi)}\right) ,
\end{equation}
and that
\begin{eqnarray}
e(\sss,1) &&= \frac{1}{\sqrt{2}}\left( [(\sss,\sss)(\sss,1)]_{(1,\sss)} +  [(\sss,\sss)(\sss,1)]_{(\psi,\sss)}\right) \nonumber\\
e(\sss,\psi)&&= \frac{1}{\sqrt{2}}\left(- [(\sss,\sss)(\sss,1)]_{(1,\sss)} + [(\sss,\sss)(\sss,1)]_{(\psi,\sss)}\right) .
\end{eqnarray}
and similarly
\begin{eqnarray}
m(\sss,1) &&= \frac{1}{\sqrt{2}}\left( [(\sss,\sss)(\sss,1)]_{(1,\sss)} -  [(\sss,\sss)(\sss,1)]_{(\psi,\sss)}\right) \nonumber\\
m(\sss,\psi) &&= \frac{1}{\sqrt{2}}\left([(\sss,\sss)(\sss,1)]_{(1,\sss)} + [(\sss,\sss)(\sss,1)]_{(\psi,\sss)}\right) .
\end{eqnarray}

We note that these are only directly read off from the gamma matrix representations. It is not yet clear if this can be understood as some generalization of fusion rules and that a corresponding set of F-matrix satisfying Pentagon relations can be defined consistent with the above identification.

Let us remark that there are more SEPs involving non-Abelian $\A$  one can construct within the framework of the twisted quantum double. A non-Abelian $\Z_2\times\Z_2\times\Z_2$ twisted gauge theory can be broken down into a novel SPT phase that involves rotation of different species. We will leave further details to a forth-coming paper\cite{HW2appear}.


\end{document}